\documentclass{PoS}

\usepackage{dsfont}
\usepackage{amsmath}
\usepackage{mathtools}
\usepackage{subcaption}

\title{Functional Fit Approach (FFA) for Density of States method: SU(3) spin system and SU(3) lattice gauge theory with static color sources}

\ShortTitle{Density of states techniques}

\author{
Mario Giuliani\textsuperscript{*}$\,$\phantom{\speaker{M. Giuliani}}\hspace*{-2.0cm}\thanks{$\;$ This work is supported by the Austrian 
Science Fund FWF, through the DK {\sl Hadrons in Vacuum, Nuclei, and Stars} (FWF DK W1203-N16) and partly by  FWF Grant I 1452-N27. Partial
support from DFG TR55, {\sl ``Hadron Properties from Lattice QCD''} is acknowledged. We thank Pascal T\"orek, Alexander Lehmann and
Axel Maas for discussions and comments on the literature, and in particular Christof Gattringer for reviewing the draft and helping in all the phases of the development of this work.}\\ \\
Universit\"at Graz, Institut f\"ur Physik, Universit\"atsplatz 5, 8010 Graz, Austria \\ \\
        E-mail: \email{mario.giuliani@uni-graz.at}      
         }

\abstract{We apply a recently developed variant of the Density of States (DoS) method, the so-called Functional Fit Approach (FFA) 
to two different models: the SU(3) spin model and SU(3) lattice gauge theory with static color sources. Both models can be 
derived from QCD and inherit the complex action problem at finite density. 
We discuss the implementation of DoS FFA in the two models and compute observables related to the particle density. 
For the SU(3) spin model we show that the results are in good agreement with the results from a Monte Carlo simulation in the dual formulation, which is free of the complex action problem. For the case of SU(3) lattice gauge theory with static color sources we 
present first results for the particle number as a function of the coupling for different values of the chemical potential.}

\FullConference{The 34th International Symposium on Lattice Field Theory\\
		24-30 July 2016\\
        University of Southampton, UK}

\begin{document}

\section{Introduction}
It is well know that a lattice simulation at finite density is, at present time, an extremely challenging task: A non-zero 
chemical potential gives rise to an imaginary part of the action which prevents the probabilistic interpretation of the Boltzmann factor, 
such that a standard MC simulation is not possible. This so-called complex action problem has stood in the way of reasonable 
progress for finite density lattice calculations, and in recent years several methods to circumvent or overcome
the complex action problem 
were proposed: Ideas such as reweighting, complex Langevin, the Lefschetz thimble, fugacity expansion, Taylor expansion, 
analytical continuation from imaginary chemical potential, or the dual formulation, are illustrating how fertile this field of research 
has been in the recent years. 

In this work we use a Density of States (DoS) method, an approach which has been reviewed also at this conference 
\cite{langfeld}, and can be used to circumvent the complex action problem. The main challenge for DoS techniques is that the 
density changes over many orders of magnitude and has to be determined with very high precision, such that it can be 
integrated over with the highly oscillating phase factor at finite chemical potential. New methods based on parameterizing the density
on small intervals and computing the parameters with restricted Monte Carlo simulations might be able to overcome the 
precision problem and could make DoS techniques a reliable tool for finite density calculations. Among these modern 
DoS techniques are the LLR approach \cite{LLR1,LLR2,LLR3,LLR4,LLR5} and the Functional Fit Approach (FFA), which we use here 
\cite{Z3_PLB,Z3_PoS,SU3_spin_NPB,SU3_spin_PoS}. For an overview of more conventional DoS strategies see, e.g., the references in the reviews \cite{langfeld,Gattringer:2016kco}.

We here discuss the implementation of DoS FFA for the SU(3) spin model, as well as for SU(3) lattice gauge theory with 
static color sources, which in the following we refer to as ''SU(3) LGT-SCS'' for brevity. 
The SU(3) spin model is a particularly suitable test case for assessing new strategies for finite density 
lattice field theory, since it has a dual representation free of the complex action problem \cite{SU3a}, and results from dual
simulations \cite{SU3b} can be used as reference data for evaluating new methods (for a complex Langevin calculation see \cite{Aarts:2011zn}). The case of SU(3) LGT-SCS 
already brings in the full dynamics of SU(3) link variables and captures the finite density 
QCD complex action problem, without having to deal with dynamical fermions (compare also the related model studied with 
DoS LLR in \cite{LLR5}).

\section{General scheme of the functional fit approach}

\subsection{Definition of the density}

\noindent
The basic equations for computing an observable $\langle O \rangle$ and the partition function $Z$ in a lattice field theory are given by
\begin{equation}
\langle O \rangle \; = \; \frac{1}{Z} \int \! D[\psi] \, e^{-S[\psi]} \, O[\psi] \quad , \qquad  Z \; = \; \int \! D[\psi] \, e^{-S[\psi]} \; .
\end{equation}
Here we use $\psi$ as a generic symbol for all fields of the theory,
and $S[\psi]$ is the corresponding Euclidean lattice action. The observable is a functional $O[\psi]$ of the field configurations $\psi$. 
By $\int D[\psi]$ we denote the lattice path integral, i.e., the product over the measures of the fields at each site (or link) of the finite lattice. 

To keep the approach as general as possible we split the action into two parts as follows,
\begin{equation}
S[\psi] \; = \; S_\rho[\psi] \; + \; \xi \, X[\psi] \; .
\label{ssplit}
\end{equation}
Both parts $S_\rho[\psi]$ and  $X[\psi]$ are real functionals of the fields $\psi$ and $\xi$ is a parameter which is either real or purely imaginary,
where the latter case is the one relevant for a situation with a complex action problem. 
Here we use a weighted density of states and $S_\rho[\psi]$ is that part of the action which we include in the density. The second part 
$X[\psi]$ is then taken into account via reweighting, 
and in case of an imaginary $\xi$, this is how the complex action problem is handled. Usually for theories with chemical potential $\mu$ the factor is of the form $\xi \propto \sinh( \mu )$.

The next step is to introduce a weighted density of states, defined as
\begin{equation}
\rho(x) \; = \; \int \! D[\psi] \, e^{-S_\rho[\psi]} \, \delta \left( X[\psi] - x \right) \; .
\end{equation} 
We can use the density to rewrite the partition function $Z$ and observables $O$, that are functions of the functional $X$: 
\begin{equation}
Z \; = \; \int_{x_{min}}^{x_{max}} \!\! dx \; \rho(x) \, e^{- \, \xi \, x} \quad , \qquad \langle O \rangle \; = \;
\frac{1}{Z}  \int_{x_{min}}^{x_{max}} \!\! dx \; \rho(x) \, e^{- \, \xi \, x} \, O[x] \; ,
\end{equation}
where $x_{min}$ and $x_{max}$ denote the bounds of $X[\psi]$. It is important to identify any symmetry of the density that can simplify the calculation. Usually it is possible to demonstrate the evenness of $\rho(x)$, using a symmetry transformation of the fields, $\psi \, \rightarrow \, \psi^{\,\prime}$, such that $S_\rho[\psi^{\,\prime}] = S_\rho[\psi]$, 
$X[\psi^{\,\prime}] = - X[\psi]$ and $\int D[\psi^{\,\prime}] = \int D[\psi]$. So we have, 
\begin{equation}
\rho(-x) \; = \; \int \! D[\psi] \, e^{-S_\rho[\psi]} \, \delta \left(X[\psi] + x \right) \; = \; 
\int \! D[\psi^{\,\prime}] \, e^{-S_\rho[\psi^{\,\prime}]} \, \delta \left(- X[\psi^{\,\prime}] + x \right) \; = \; \rho(x) \; .
\end{equation}
In that case the expressions for $Z$ and $\langle O \rangle$ simplify to
\begin{equation}
Z \, = \, 2 \int_0^{x_{max}} \!\! dx \; \rho(x) \, \cosh( \xi \, x) \; , \; \langle O \rangle \, = \,
\frac{2}{Z}  \int_0^{x_{max}} \!\! dx \; \rho(x) \, \Big( \!\cosh( \xi \, x) \, O_e [x] -  \sinh( \xi \, x) \, O_o [x] \!\Big)  \; ,
\label{evenobs}
\end{equation}
where we have defined the even and odd parts of the observable as $O_e[x] = (O[x] + O[-x])/2$, $O_o[x] = (O[x] - O[-x])/2$.
Eq.~(\ref{evenobs}) shows that we have halved the initial work, because only the density for positive $x$ is needed. We further observe that for purely imaginary $\xi$ the hyperbolic functions are replaced by trigonometric ones, $\cos( \xi \, x)$ and $i\sin( \xi \, x)$. They are strongly oscillating functions and this is how the sign problem shows itself in the DoS approach. 

\subsection{Parametrization of the density}

\noindent 
The next step is to find a suitable parametrization of the density $\rho(x)$, which in many cases  
is an exponential ansatz which we discuss now. For the parametrization we first divide the range $[0,x_{max}]$
into $N$ intervals $I_n=[x_n,x_{n+1}],n=0,1,\dots,N-1$ of variable size $\Delta_n = x_{n+1}-x_{n}$ and $\sum_{n=0}^{N-1} \Delta_n = x_{max}$. 
The exponential ansatz for $\rho(x)$ now has the form 
\begin{equation}
\rho(x) \; = \; e^{\, -l(x)} \qquad \mbox{with} \qquad  l(x) \; = \; d_n \, + \, x \, k_n \quad \mbox{for} \quad x \, \in \, I_n = [x_n, x_{n+1}] \; .
\label{rhoparam}
\end{equation}
In other words the density is the exponential of a piecewise linear function $l(x)$. For each interval this linear function is
parameterized by a constant $d_n$ and a slope $k_n$. We furthermore require $l(x)$ to be continuous and normalize it to $l(0) = 0$, which 
corresponds to the normalization $\rho(0) = 1$ (such a normalization can be chosen freely). The continuity condition and the normalization 
completely determine the constants $d_n$, and the slopes $k_n, \, n = 0,1 \, ... \, N-1$ are the parameters of $\rho(x)$.

\subsection{Determination of the parameters with restricted expectation values}

\noindent 
For the determination of the parameters $k_n$ we define so-called restricted expectation values 
$\langle\!\langle O \rangle\!\rangle_n(\lambda)$. They depend on a real parameter $\lambda$ and are defined as:
\begin{eqnarray}
\langle\!\langle O \rangle\!\rangle_n(\lambda) & = & \frac{1}{Z_n(\lambda)} \int \! D[\psi] \, e^{\, -S_\rho[\psi] \, + \, \lambda X[\psi]} \, 
O[X[\psi]] \, \Theta_n[X[\psi]] \; , 
\nonumber \\
Z_n(\lambda) & = & \int \! D[\psi] \, e^{\, -S_\rho[\psi] \, + \, \lambda X[\psi]} \, \Theta_n[X[\psi]] \; ,
\nonumber \\
\Theta_n[x] & = & \left\{ 
\begin{array}{cl}
1 & \; \mbox{for} \; x \, \in \, [x_n, x_{n+1}]  \\
0 & \quad \mbox{otherwise}
\end{array} \right. \; .
\label{restrictvev}
\end{eqnarray}
The support function $\Theta_n[x]$ restricts the simulation to the interval $I_n$ that we want to sample, where the slope $k_n$ determines $\rho(x)$. The additional Boltzmann weight with the parameter $\lambda \in \mathbb{R}$ can be used to systematically probe the density in $I_n$. Since $\lambda$ is real, one can use a standard simulation to compute the restricted expectation value. 

One can now work out explicitly $Z_n(\lambda)$ using the parametrized density of states. One finds
\begin{equation}
Z_n(\lambda) \; = \; \int_{x_n}^{x_{n+1}} dx \, \rho(x) \, e^{\, \lambda \, x} \; = \text{c} \! \int\limits_{x_n}^{x_{n+1}}dx \, e^{- k_n x} \, e^{\lambda x}  \\ = c \frac{ e^{( \lambda - k_n )x_{n+1}} - e^{( \lambda - k_n ) x_{n} } }{ \lambda - k_n } \; ,
\label{onrho}
\end{equation}
where $c$ is an irrelevant constant. From this we can compute the expectation value $\langle\!\langle X \rangle\!\rangle_n(\lambda)$, which we write
as $\langle\!\langle X \rangle\!\rangle_n(\lambda) \; = \; \partial \ln Z_n(\lambda) / \partial \lambda$.
A straightforward calculation gives
\begin{equation}
\frac{1}{\Delta_n} \Big( \langle\!\langle X \rangle\!\rangle_n(\lambda) - x_n \Big) \, - \, \frac{1}{2} \;\; = \;\; F\big( (\lambda - k_n) \Delta_n \big) \quad \text{with} \quad F(s) \; = \; \frac{1}{1 - e^{-s}} \, - \, \frac{1}{s} \, - \, \frac{1}{2} \; .
\label{fitfunction}
\end{equation}
$F(s)$ is a real-valued smooth function of a real variable $s$. It is monotonically increasing and has a single zero at $s=0$. $F\big( (\lambda - k_n) \Delta_n \big)$ describes the distribution of the restricted expectation values in the l.h.s of Formula\;(\ref{fitfunction}) as a function of $\lambda$. Performing M.C. simulations for different values of $\lambda$ gives us a set of points with an associated error. We fit these points with $F \big( (\lambda - k_n) \Delta_n \big)$ and we obtain the slope $k_n$. This procedure is the reason for referring to our method as the "functional fit approach". Once the $k_n$ are computed we know the density and obtain observables using Eq.\;(\ref{evenobs}).

\section{The SU(3) spin system and SU(3) lattice gauge theory with static color sources}
\label{su3}

The SU(3) spin model and the SU(3) LGT-SCS are two closely related models. Both can be motivated from QCD and inherit the complex action problem from QCD. Results for the SU(3) spin model were already discussed in \cite{SU3_spin_NPB}, such that here the discussion of that case is kept short.

For the SU(3) spin model the action is,
\begin{equation}
S[ \mathrm{P} ] \; = \; - \tau \, \sum\limits_{ \vec{n} \in \Lambda_3 } \sum\limits_{\nu=1}^3 
\Bigl[  \mathrm{P}( \vec{n} ) \, \mathrm{P}(\vec{n}+\hat\nu)^{*} + c.c. \Bigr]
 \; - \; \kappa\,  \sum\limits_{ \vec{n} \in \Lambda_3 } \Bigl[ e^\mu  \, \mathrm{P}(\vec{n}) + e^{-\mu} \, \mathrm{P}(\vec{n})^{*} \Bigr] \; ,
\label{actio-SU3-spin}
\end{equation}
where the $\mathrm{P}( \vec{n} )$ are the traces of SU(3) matrices on the sites $\vec{n}$ of a 3-dimensional lattice representing the Polyakov loops.

For SU(3) LGT-SCS the action is
\begin{equation}
S[U] = - \frac{ \beta }{3} \sum\limits_{ n \in \Lambda } \sum\limits_{ \mu < \nu } \mathrm{Re} \bigl[ \mathrm{Tr} ( U_{ \mu }(n) \, U_{ \nu }(n + \hat\mu) \, U_{ \mu }(n + \hat\nu)^{\dagger} U_{ \nu }(n)^{\dagger} ) \bigl] - \kappa \sum\limits_{ \vec{ n } \in \Lambda_3 } \bigl[ e^{ \mu N_T } \mathrm{P}( \vec{ n } ) + e^{ - \mu N_T } \mathrm{P}( \vec{ n } )^{\dagger} \bigl] \; ,     
\label{actio-SU3-static-quarks}
\end{equation}
where here the d.o.f. are SU(3) matrices $U_{\mu}(n)$ on the links of a 4-dimensional lattice, and we have a proper Polyakov loop defined as:

\begin{equation*}
\mathrm{P}(\vec{n})= \frac{1}{3} \mathrm{Tr} \prod^{N_T-1}_{n_4=0} \, \mathrm{U}_4( \vec{n} , n_4 ) \; ,
\end{equation*}
and the corresponding terms in (\ref{actio-SU3-static-quarks}) represent static color sources.

The two models have the parameter $\kappa$ in common, which is a decreasing function of the quark mass. Also the chemical potential has the same role in the two theories: it gives a different weight to forward and backward winding Polyakov loops. Note that in the SU(3) LGT-SCS we display explicitly the temporal extent $N_T$, while in the spin model this is absorbed in the chemical potential $\mu$.
The two models differ in the terms describing the dynamics of the degrees of freedom: In the SU(3) LGT-SCS we use the Wilson gauge action with inverse gauge coupling $\beta$, while in the spin model this is replaced by a nearest neighbour interaction and $\tau$ is the temperature parameter.

The decomposition of the action is quite straightforward for both models. We can write:
\begin{equation}
\begin{split}
& S_\rho[ \psi ] \; = \; \mbox{Re} \, S[ \psi ] \; , \; \; \\
& \text{SU(3) spin} : \, X[ \mathrm{P} ] \; =  \; \sum\limits_{ \vec{n} } \mbox{Im} \, \mathrm{P}( \vec{ n } ) \; , \; \; \xi \; = \; i \, 2 \kappa \sinh(\mu) \; , \\
& \text{SU(3) LGT-SCS} : \, X[ \mathrm{U} ] \; =  \; \sum\limits_{ \vec{n} } \mbox{Im} \, \mathrm{P}( \vec{ n } ) \; , \; \; \xi \; = \; i \, 2 \kappa \sinh(\mu N_T) \; . \\
\end{split}
\end{equation}  
The symmetry which ensures the evenness of the density $\rho(x)$ is complex conjugation of the dynamical variables in both models.

With these definitions it is simple to implement the FFA and to perform the required restricted Monte Carlo simulations for finding the slopes $k_n$ which determine the density $\rho(x)$. We can define the same observables for the two models by derivatives with respect to $\sinh(\mu$) (for SU(3) spin) and with respect to $\sinh(\mu N_T)$ (for SU(3) LGT-SCS):
\begin{equation}
n \;\; \equiv \; \langle \, \mbox{Im} \; \mathrm{P}( \vec{ n } ) \, \rangle \, \equiv \;\; - \frac{1}{V}\frac{1}{2\kappa} \frac{\partial}{\partial \sinh ( \mu )} \, \ln Z \;\; = \;\; \frac{2}{V} 
\frac{1}{Z} \int\limits_{0}^{x_{max}} \!\!\!  dx \; \rho(x) \, \sin( 2\kappa \sinh ( \mu ) x ) \, x \; ,
\end{equation}
\begin{equation}
\begin{split}
\chi_{n} & = \frac{1}{2\kappa} \, \frac{\partial}{\partial \sinh( \mu ) } n \\ & = \frac{1}{V} \, \biggl\{ \frac{2}{Z} \int\limits_{0}^{x_{max}} dx \, \rho(x) \, \cos(2 \kappa \sinh( \mu ) x) \, x^2 + \Bigl( \frac{2}{Z} \int\limits_{0}^{x_{max}} dx \, \rho(x) \, \sin( 2 \kappa \sinh( \mu ) x \Bigr)^2 \biggr\} ~,
\end{split}
\label{eq:susce_n}
\end{equation}
where we obtain the corresponding observables for SU(3) LGT-SCS by replacing $\mu$ with $\mu N_T$.

\begin{figure}
\begin{subfigure}{.5\textwidth}
  \hspace{-0.6cm}
  \includegraphics[width=1.075\textwidth]{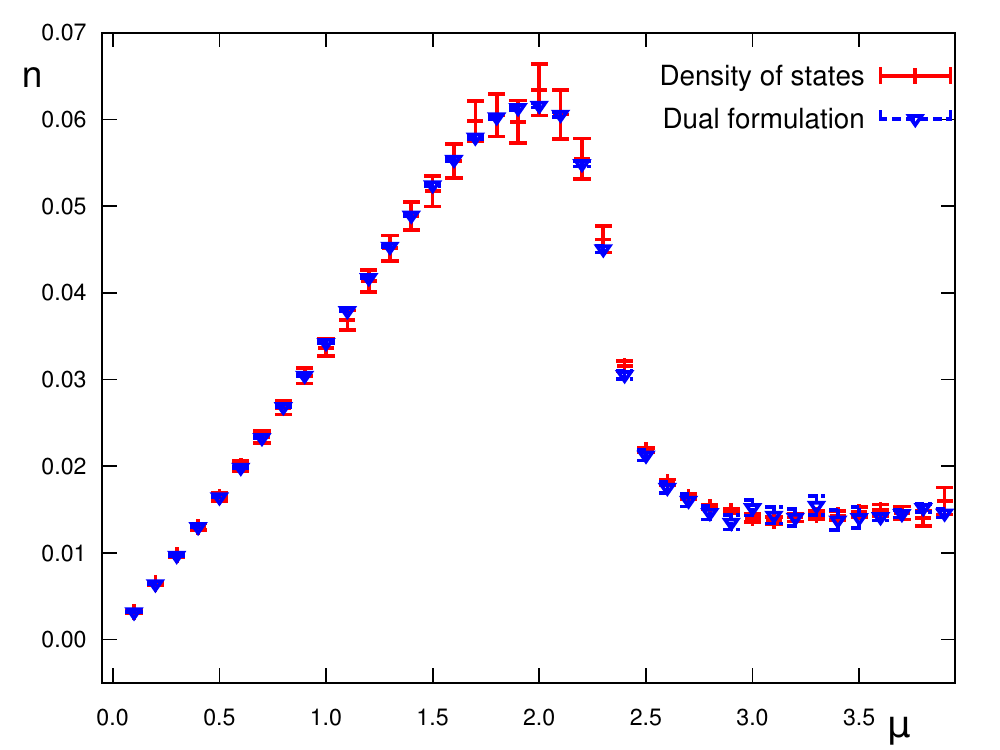}
\end{subfigure}%
\begin{subfigure}{.5\textwidth}
  \centering
  \includegraphics[width=1.075\textwidth]{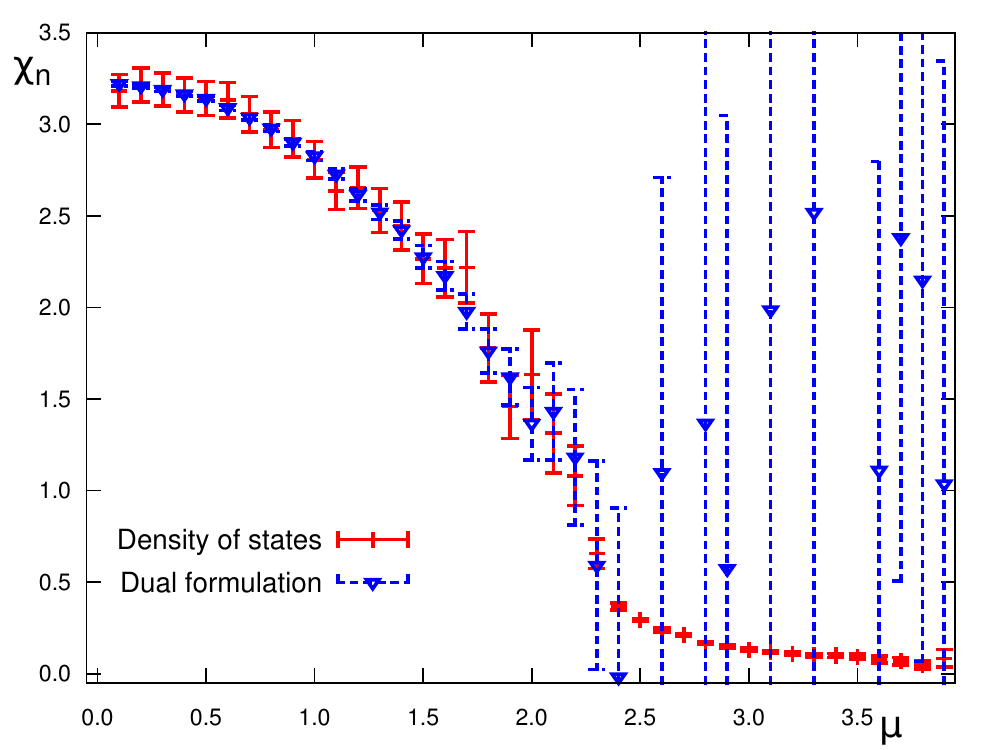}
\end{subfigure}
\caption{Simulation of the SU(3) spin model. Observable for a lattice $12^3$ with $\kappa=0.005$ and $\tau=0.130$ as function of $\mu$. We observe a good agreement between the two methods.}
\label{fig:results_12_murun}
\end{figure}

Let us now come to the presentation of the results for these observables. As already remarked, for the SU(3) spin system we can use the results from the dual formulation \cite{SU3a,SU3b} as reference data for the DoS FFA results. In Fig.\,\ref{fig:results_12_murun} we show $n$ and $\chi_n$ as a function of $\mu$ for a $12^3$ lattice at fixed $\tau=0.130$ and $\kappa=0.005$. We observe good agreement between the two different methods up to $\mu \approx 4$ for $n$ and up to $\mu \approx 2$ for $\chi_n$. This confirms the reliability of the DoS FFA for $n$ and $\chi_n$ in the respective ranges of $\mu$( see \cite{SU3_spin_NPB} for a more detailed discussion).

\begin{figure}
\begin{subfigure}{.5\textwidth}
  \hspace{-0.8cm}
  \includegraphics[width=1.075\textwidth]{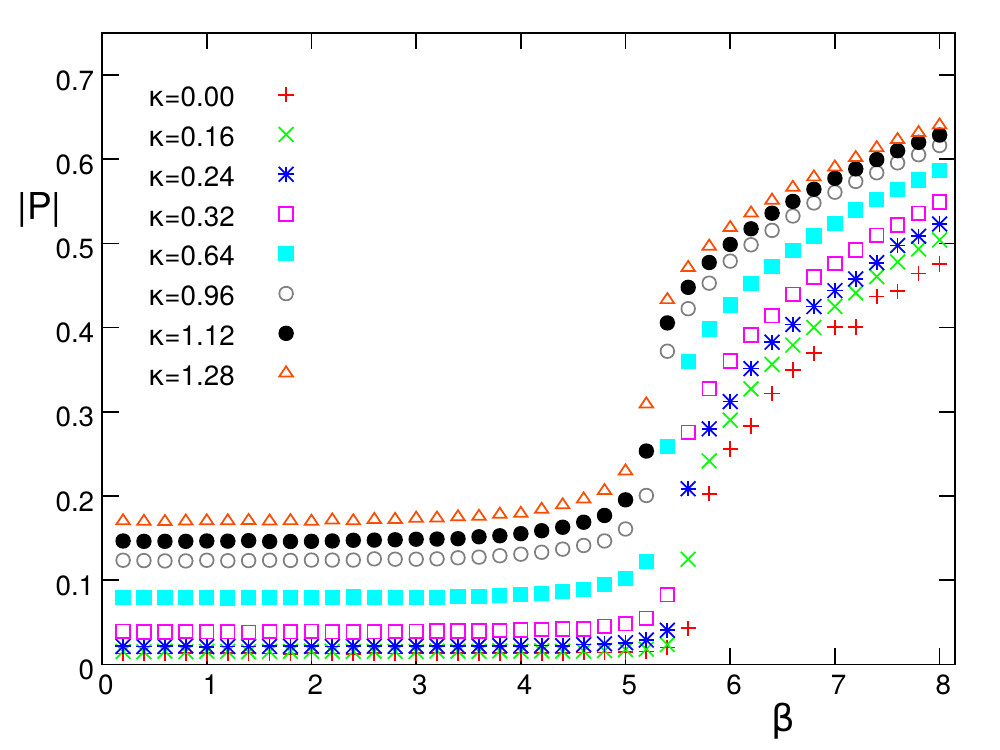}
\end{subfigure}%
\begin{subfigure}{.5\textwidth}
  \centering
  \includegraphics[width=1.075\textwidth]{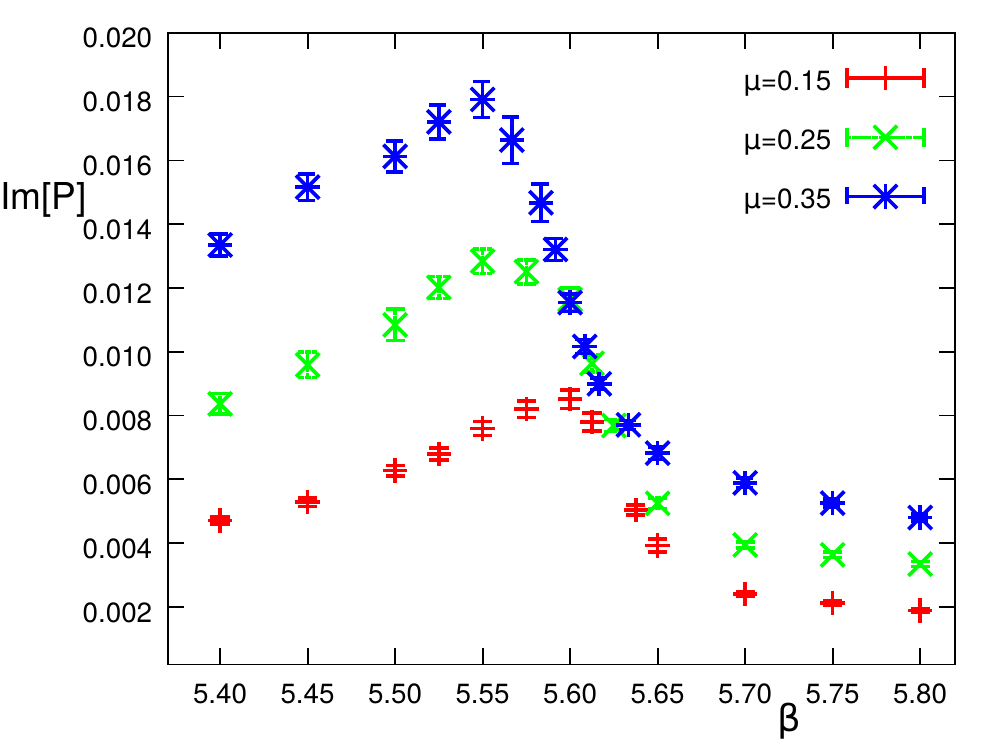}
\end{subfigure}
\caption{Results for SU(3) LGT-SCS on a $8^3\times4$ lattice. On the l.h.s.\,we present $\langle | \mathrm{P} | \rangle $ for a standard simulation at $\mu=0$. We observe a shift toward smaller $\beta$ for larger $\kappa$. On the r.h.s.\,we show $n=\langle \mathrm{Im \, P} \rangle$ from a simulation at $\mu \neq 0$ and fixed $\kappa=0.04$. We observe a shift toward smaller $\beta$ for increasing chemical potential $\mu$.}
\label{fig:standard_static_quarks}
\end{figure}

SU(3) LGT-SCS, up to today, has no dual formulation. To study this model we decided to perform a preliminary standard simulation at $\mu=0$. This is useful because it allows to study the transition as function of $\kappa$. In the l.h.s.\,plot of Fig.\,\ref{fig:standard_static_quarks} we show $\langle | \mathrm{P} | \rangle $ as a function of $\kappa$ and observe that for larger $\kappa$ the transition shifts toward smaller $\beta$. Next we try to locate the bending of the surface also at chemical potential $\mu \neq 0$. Pursuing this goal we simulated a $8^3\times4$ lattice with $\kappa=0.04$ and different $\beta$ and chemical potentials. In the r.h.s.\,plot of Fig.\,\ref{fig:standard_static_quarks} we show $ n = \langle \mathrm{Im \, P} \rangle $ as a function of $\beta$ and observe a shift toward smaller $\beta$ for larger chemical potential $\mu$. A detailed discussion of this model will be presented in a forthcoming paper.

\section{Summary}
\label{summary}
In these proceedings we briefly reviewed the Functional Fit Approach for the DoS method.  This method is a general tool to handle the complex action problem. The argument of the density $\rho$ is divided in a set of intervals and $\rho$ is parameterized as the exponential of a piecewise linear function. In this way $\rho$ is completely described by the slopes $k_n$ of the intervals $\Delta_n$, that we can compute using a restricted Monte Carlo simulation.

We briefly present two models, SU(3) spin and SU(3) lattice gauge theory with static color sources. The results are encouraging and we conclude that the FFA approach is an interesting tool to study systems with a complex action problem.

\end{document}